\documentclass[preprint]{aastex}
\singlespace
\def\p/{\mbox{$^1$}}
\def\pp/{\mbox{$^2$}}
\def\ppp/{\mbox{$^3$}}
\def\pppp/{\mbox{$^4$}}
\def\m/{\mbox{$^{-1}$}}
\def\mm/{\mbox{$^{-2}$}}
\def\mmm/{\mbox{$^{-3}$}}
\def\mmmm/{\mbox{$^{-4}$}}
\def\zsun/{\mbox{$z_\sun$}}
\def\pio{\mbox{$\pi_{\rm o}$}}
\def\ro{\mbox{$r_{\rm o}$}}
\def\rhoo{\mbox{$\rho_{\rm o}$}}
\def\sigmastard{\mbox{$\sigma_{\star,{\rm disk}}$}}
\def\sigmastarh{\mbox{$\sigma_{\star,{\rm halo}}$}}
\newcommand{\sci}[2]{\mbox{$#1\cdot 10^{#2}$}}

\slugcomment{To appear in the May 2001 issue of the Astronomical Journal}
\lefthead{Ma\'{\i}z-Apell\'aniz  2001}
\righthead{The spatial distribution of O-B5 stars}
\begin{document}

\title{The spatial distribution of O-B5 stars in the solar 
neighborhood as measured by {\em Hipparcos}\altaffilmark{1}}
\shorttitle{The spatial distribution of O-B5 stars}

\author{Jes\'us Ma\'{\i}z-Apell\'aniz}
\affil{Space Telescope Science Institute\altaffilmark{2}, 3700 San Martin 
Drive, Baltimore, MD 21218, U.S.A.}


\altaffiltext{1}{Based on data from the {\em Hipparcos} astrometry
satellite.}
\altaffiltext{2}{The Space Telescope Science Institute is operated by the
Association of Universities for Research in Astronomy, Inc. under NASA
contract No. NAS5-26555.}

\begin{abstract}

	We have developed a method to calculate the fundamental parameters of 
the vertical structure of the Galaxy in the solar neighborhood from
trigonometric parallaxes alone. The method takes into account Lutz-Kelker-type 
biases in a self-consistent way and has been applied to a sample of O-B5
stars obtained from the Hipparcos catalog. We find that the Sun is 
located 24.2~$\pm$~1.7~(random)~$\pm$~0.4~(systematic)~pc above the galactic 
plane and that the disk O-B5 stellar population is distributed with a scale 
height of 34.2~$\pm$~0.8~(random)~$\pm$~2.5~(systematic)~pc and an integrated 
surface density of 
(1.62~$\pm$~0.04~(random)~$\pm$~0.14~(systematic)$)\cdot 10^{-3}$ stars pc\mm/. 
A halo component is also detected in the distribution and constitutes at least
$\approx 5$\% of the total O-B5 population.
The O-B5 stellar population within $\sim 100$ pc of the Sun has an anomalous 
spatial distribution, with a less-than-average number density. This local 
disturbance is probably associated with the expansion of Gould's belt.

\end{abstract}


\section{INTRODUCTION}

	Studies based on the spatial distribution of massive stars 
\citep{Blaa60} and interstellar hydrogen \citep{Gumetal60} established some time
ago that the Sun is located above the galactic plane. This has been confirmed
using a variety of tracers (see \citealt{HumpLars95} for a review) with somewhat
incompatible values of \zsun/, the Sun's distance above the galactic plane, 
ranging from 9.5 pc to 42 pc. Some of the discrepancies may be due to real 
differences in the vertical distribution of
different stellar or gaseous components but others must be due to observational
errors, since they appear among studies of the same type of objects. This is
seen in Table~\ref{previous}, where the results on early-type stars are shown. 
The use of early-type stars to determine \zsun/ has the advantage that their
scale height with respect to the galactic plane, $h$, is of the same order of 
magnitude as \zsun/, which makes the determination simpler than for objects with
larger scale heights (e.g. late-type stars). However, early-type stars are 
scarce and to obtain a statistically-reliable sample previous studies had to
resort to objects close to the galactic plane, where extinction can introduce 
severe biases.

	The measurement of \zsun/ using early-type stars is closely related to
the vertical distribution of the number density of those stars, 
$\rho(z)$. We consider here two scenarios. First, if the analyzed sample
consists of a single population which is responsible for the gravity field in 
which the stars are moving we have a
a single component, self-gravitating, isothermal disk (e.g. \citealt{King96}),
which can be described by:

\begin{equation}
\rho(z) = \rhoo\, {\rm sech}^2 \left(\frac{z+\zsun/}{2h_s}\right), \label{dens1}
\end{equation}

\noindent where $z$ is the vertical coordinate (with its origin at the Sun's
position), $h_s$ is the corresponding scale height, and $\rhoo$ is the density 
at the galactic plane. The factor of 2 is introduced in order to have 
$\rho \propto \exp(-|z+\zsun/|/h_s)$ for $|z+\zsun/| \gg h_s$. A 
second possible scenario can be introduced by assuming the same conditions as
before but now establishing that the gravity field is caused by a different 
population (e.g. low-mass stars or dark matter) with constant density. 
In that case we have a Gaussian vertical distribution:

\begin{equation}
\rho(z) = \rhoo\, \exp\left(-\frac{1}{2}\left(\frac{z+\zsun/}
{h_g}\right)^2\right), 
\label{dens2}
\end{equation}

\noindent where $h_g$ is the Gaussian half-width.

	The stellar surface density, $\sigma_\star$, can be obtained by 
integrating $\rho(z)$ from $-\infty$ to $+\infty$ and for the models described 
by Eqs.~\ref{dens1}~and~\ref{dens2} the results are $4h_s\rhoo$ and 
$\sqrt{2\pi}h_g\rhoo$, respectively. Measured values of $h$ and 
$\sigma_\star$ have been included in Table~\ref{previous}. In each case it is 
indicated whether the model used is that of Eq.~\ref{dens1} or an exponential 
one: $\rho(z) = \rhoo\, \exp (-|z+\zsun/|/h_e)$, in which case 
$\sigma_\star = 2h_e\rhoo$.

\begin{deluxetable}{llllcclc}
\tablecaption{Previous determinations of parameters associated with the 
vertical distribution of the early-type stellar population. The last column was
calculated by multiplying the measured surface density by a factor derived from
an unbiased O+B sample.\label{previous}}
\tabletypesize{\scriptsize}
\tablewidth{0pt}
\tablehead{\colhead{Source} & \colhead{$h$} & \colhead{\zsun/} & 
\colhead{Range\tablenotemark{a}} & \colhead{Model\tablenotemark{b}} & 
\colhead{$\sigma_{\star,\,\rm meas}$\tablenotemark{c}} & 
\colhead{Range\tablenotemark{d}} & 
\colhead{$\sigma_{\star,\,\rm equiv}$\tablenotemark{e}} \\
 & \colhead{(pc)} & \colhead{(pc)} & & &
\colhead{(stars pc\mm/)} & & \colhead{(stars pc\mm/)} }
\startdata
\citet{Alle73}        & 60                         & \colhead{-----} & 
O+B             & exponential      & \sci{1.2}{-2}       & O+B              & 
\sci{2.6}{-3}         \\
\citet{StotFrog74}    & $46\pm 7$\tablenotemark{f} & $\;24\pm 3$     & 
O-B5            & exponential      & \sci{5.3}{-4}       & O-B3.5           &
\sci{0.85}{-3}         \\
                      & $70\pm 3$\tablenotemark{g} &                 &
                &                  &                     &                  &
                      \\
\citet{ContVacc90}    & $45\pm 5$                  & $\;15\pm 3$     &
WR              & self-gravitating & -----               & \colhead{-----}  & 
-----                 \\
\citet{Reed97,Reed00} & $45\pm 20$                 & $9.5\pm 3.5$    &
O-B2            & exponential      & \sci{(0.6-8.0)}{-4} & O-B2             & 
\sci{(0.15-1.95)}{-3} \\
\enddata
\tablenotetext{a}{Spectral subtypes used for the scale height and Sun's distance
above the galactic plane determinations.} 
\tablenotetext{b}{Model used for the scale height determination.} 
\tablenotetext{c}{Measured surface density.} 
\tablenotetext{d}{Spectral subtypes used for the surface density 
determination.} 
\tablenotetext{e}{Equivalent surface density for O-B5 stars.} 
\tablenotetext{f}{Within 200 pc of the Sun.} 
\tablenotetext{g}{Within 800 pc of the Sun.} 
\end{deluxetable}

	The profiles defined by Eqs.~\ref{dens1}~or~\ref{dens2} are actually
not extremely different from one another and either one of them is expected to
be a reasonably good approximation to the vertical profile of the disk
population of early-type stars. However, we must bear in mind that there are 
also early-type stars in the halo (see, e.g. \citealt{Rolletal99} and references
therein) which are expected to dominate the O-B5 population at distances larger
than several hundred pc from the galactic plane and, thus, add another
component to the observed disk population. Most of these objects are 
runaway stars \citep{Rolletal99,Hoogetal00} but some appear to have been 
formed in situ \citep{Conletal92}.

	The availability of the Hipparcos catalog \citep{ESA97} has 
revolutionized the quality of astrometric data and has produced for the first 
time good-quality trigonometric parallaxes for a large sample of early-type
stars. However, even with these data one has to take into account a
fundamental problem in the use of trigonometric parallaxes for the determination
of distances discovered by \citet{LutzKelk73}. Assuming a uniform 
spatial distribution of stars and a Gaussian distribution of observed 
parallaxes $\pio$ about the true parallax $\pi$ with a width $\sigma_\pi$, 
they discovered that the distribution of true distances $r = 1/\pi$ about the 
observed one $\ro = 1/\pio$ was asymmetrical and even divergent for 
$r\rightarrow\infty$. The effect is caused by the inverse relationship between 
$\pi$ and $r$ and by the fact that the available volume between $r$ and $r+dr$
is proportional to $r^2$. The effect is unimportant for parallaxes with 
small relative errors ($\varepsilon_\pi \equiv \sigma_\pi/\pio \lesssim 0.05$) 
and can be corrected for cases with $\varepsilon_\pi \approx 0.10-0.15$ but for 
values with $\varepsilon_\pi \gtrsim 0.175$ it becomes so large
that an individual trigonometric parallax by itself provides little information 
about the distance to the object unless it is combined with additional 
information (e.g. the spectroscopic parallax). \citet{LutzKelk73} provided
a table to correct the effect of the bias for $\varepsilon_\pi < 0.175$ (see 
\citet{Smit87,Kova98} for updated analyses) but for larger values no simple
meaningful correction exists. 

	A large fraction of 
the Hipparcos trigonometric parallaxes have $\varepsilon_\pi > 0.175$ and a 
non-negligible number have negative parallaxes. Those values cannot be 
discarded when statistical studies of distances are undertaken without 
introducing severe biases. One approach to the problem is that of 
\citet{RatnCase91}, who use spectroscopic parallaxes to constrain the possible 
range of distances without discarding any stars with large values of 
$\varepsilon_\pi$ or negative ones of $\pio$. Another approach is 
that followed by \cite{FeasCatc97} to evaluate the zero-point of the Cepheid
period-luminosity relation. Those authors combine trigonometric parallaxes with
$\varepsilon_\pi$ up to $\approx 1.0$ with the apparent magnitude and period of 
the stars to reduce all Cepheids to a single distance and produce an equivalent 
parallax with $\varepsilon_\pi \approx 0.05$.

	In this paper we introduce a new method to correct for Lutz-Kelker-type
biases (section 2) and apply it to a sample of stars obtained from the 
Hipparcos catalog (section 3) in order to derive the parameters which
determine the vertical distribution of O-B5 stars in the solar neighborhood 
(section 4).

\section{DESCRIPTION OF THE METHOD}

	The number of stars per unit distance $n_{\star,r}$ between $r$ and 
$r+dr$ inside a solid angle $\Omega$, $n_{\star,r} dr$, is given by:

\begin{equation}
n_{\star,r} dr = \rho(r)\, \Omega\, r^2\,dr, \label{nstar1}
\end{equation}

\noindent where $\rho(r)$ is the star number density. The equivalent
expression per unit parallax is:

\begin{equation}
n_{\star,\pi} d\pi = \rho(1/\pi)\, \Omega\, \pi\mmmm/\,d\pi. \label{nstar2}
\end{equation}

	If we assume that the distribution of $\pio$ about $\pi$ is Gaussian
with a width $\sigma_\pi$, then the distribution of observed parallaxes 
will be the convolution of the two functions:

\begin{equation}
f(\pio) = \int_{0}^{\infty} 
A\, e^{-\frac{1}{2}\left(\frac{\pi-\pi_{\rm o}}{\sigma_\pi}\right)^2} 
\rho(1/\pi)\, \Omega\, \pi\mmmm/\,d\pi, \label{dist1}
\end{equation}

\noindent where $A$ is a normalizing factor. Note that Eq.~\ref{nstar2} may be
singular at $\pi = 0$ for some choices of $\rho(r)$ and that this may cause 
Eq.~\ref{dist1} to be non-normalizable. This is the reason why no simple 
correction can be obtained for the standard case of Lutz-Kelker bias (which has
$\rho(r) = \rhoo$) when $\varepsilon_\pi \gtrsim 0.175$. A generalization of 
that standard case can be achieved by allowing for non-constant density 
profiles, as previously suggested by \citet{Hans79}. 
This is motivated by the finite size of the Galaxy, which forces the
introduction of a cutoff in any choice of $\rho(r)$ based on a realistic model.
Such a cutoff would eliminate any convergence problems in Eq.~\ref{dist1} and
allow for finite Lutz-Kelker-type corrections\footnote{Note, however, that such
corrections will depend on $\rho(r)$ itself and that, for a non-spherically
symmetric density distribution (as expected for a galactic disk), $\rho(r)$ will
be different for each direction. Thus, a tabulated version of Lutz-Kelker-type
corrections becomes clearly impractical due to the increase in the number of
degrees of freedom and the best strategy to estimate the distance is to compute 
an individual correction for each star.}. 

	We consider one such model based on a self-gravitating isothermal
vertical distribution (Eq.~\ref{dens1}) and which for early-type stars should 
be valid for large absolute values of the galactic latitude $b$, where the main 
limitation imposed on sample completeness is distance (as opposed to 
extinction): 

\begin{equation}
\rho(1/\pi) = \rhoo\, {\rm sech}^2\left(\frac{\sin b/\pi+\zsun/}{2h_s}\right) =
     \rhoo\, {\rm sech}^2\left(\frac{1/\pi\pm\zsun/^\prime}{2h_s^\prime}\right),
\label{dens3}
\end{equation}

\noindent where $h_s^\prime = h_s\,|\csc b|$ and 
\zsun/$^\prime = \zsun/\,|\csc b|$, and the sign in the last expression is 
positive for $b>0$ and negative for $b<0$. 
A least-squares fit to the observed distribution of parallaxes using
Eqs.~\ref{dist1}~and~\ref{dens3} should then yield the three free 
parameters $h_s$, \zsun/, and $\sigma_\star$ (which is a function of
the constants in the two equations). Unfortunately, two problems exist when this
is attempted. First, a large range in $b$ needs to be used in order to have a
large enough sample, which implies that a mean value of $|\csc b|$ will have to
be obtained in order to derive $h_s$ and \zsun/ from $h_s^\prime$ and 
\zsun/$^\prime$. However, for $|b| \lesssim 25\degr$ and reasonable values of 
$h_s$ and \zsun/, $|\csc b|$
becomes so large that we are forced to go to distances larger than 1 kpc in
order to have reasonably complete samples, and this is beyond the volume
well-sampled by Hipparcos. Therefore, this method cannot be applied to the data
currently available. The second problem is more subtle: Suppose we analyze all
the stars contained within a double cone defined by $|b| > b_{\rm min}$ and that
the vertical spatial distribution of early-type stars is given by the sum of
either Eq.~\ref{dens1} or Eq.~\ref{dens2} and a halo component which is
unimportant near the galactic plane but dominates at high $|z|$. Then, since 
the sampled volume is proportional to $z^2$, the halo component will be
overrepresented in our sample and a strong bias could be introduced in our 
measurements of $h_s$ and \zsun/ unless the proper linear combination of
$\rho_{\rm disk}$ and $\rho_{\rm halo}$ is used.

%
%

	An alternative method which avoids these two problems
can be designed in the following way. Suppose we can make
an educated guess for the parameters that describe the O-B5 disk population 
(using either Eq.~\ref{dens1} or Eq.~\ref{dens2}) and that we also have a 
reasonably good approximation for the halo population (which is expected to be 
only a small contaminant in our sample if we avoid the bias mentioned in the
previous paragraph). Then, for a single star with measured 
$\pio$, $\sigma_\pi$, and $b$ we can calculate the probability distribution as 
a function of distance as:

\begin{equation}
p(r) = A\, r^2\,e^{-\frac{1}{2}\left(\frac{1-r\pi_{\rm o}}
{r\sigma_\pi}\right)^2} \rho(r), \label{dist2}
\end{equation}

\noindent where $A$ is a normalizing factor and the latitude dependence is
embedded in $\rho(r) = \rho_{\rm disk}(r) + \rho_{\rm halo}(r)$. We can now 
project $p(r)$ into the vertical coordinate to obtain $p(z)$ and sum over all 
observed stars to obtain $f(z)$, the distribution of stars as a function of 
the vertical coordinate. Doing a least-squares fit of $f(z)$ with a vertical 
profile leads to a new estimate of the parameters. The process is then repeated 
until convergence is achieved.

\begin{figure}
\centerline{\includegraphics*[width=\linewidth]{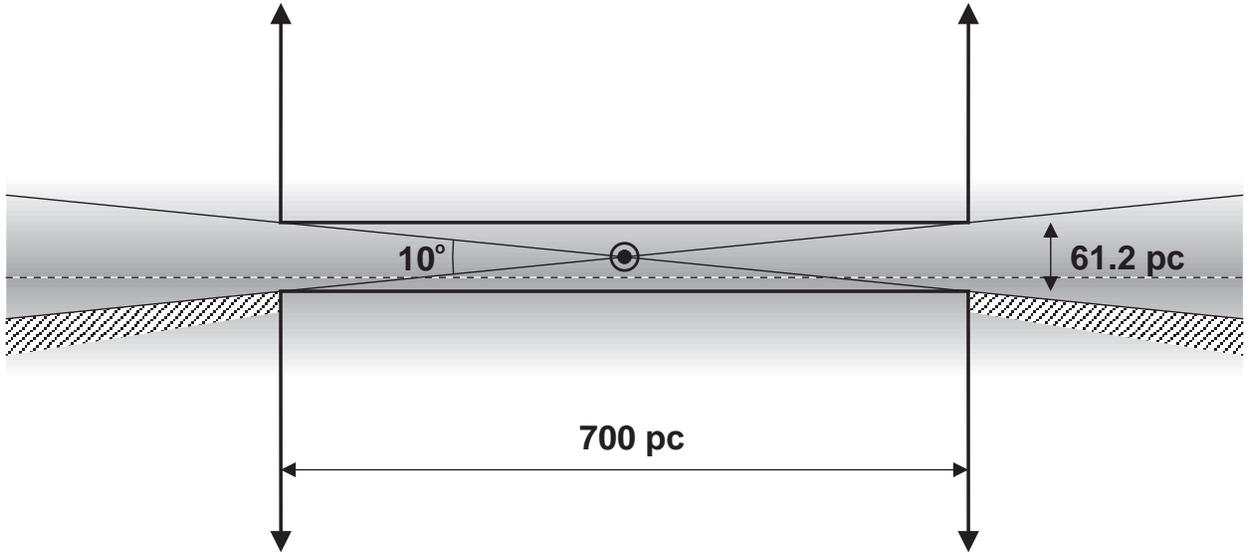}}
\caption{The double semi-infinite cylinder used in this paper to study the 
vertical distribution of O-B5 stars is represented by the two regions 
contained inside the thick lines. The values of the parameters used are
$\varpi_{\rm max} = 350$ pc and $b_{\rm min} = 5\degr$. The shaded area 
represents the galactic disk with the dashed line being the galactic plane and 
the Sun symbol marking the position of our star. The hatched regions mark the 
area excluded to estimate the bias induced by extinction (see the text for 
details).}
\label{cylmodel}
\end{figure}

	Several precautions have to be taken to insure the correctness of this
iterative process. First, a maximum distance to the Sun vertical line,
$\varpi_{\rm max}$, has to be specified in order to insure completeness and to
avoid the halo-disk bias previuosly discussed. This can
be achieved by truncating each individual $f(z)$ at 
$z_{\rm max} = \varpi_{\rm max}\tan b$. Second, stars at very low galactic
latitudes ($|b| < b_{\rm min}$) have to be excluded from the sample since for
lines of sight close to the galactic equator the spatial structure is dominated
by radial (not vertical) variations and extinction. In order not to induce any
biases due to this exclusion, the fit to the vertical profile should not include
the range 
$(-\varpi_{\rm max}\tan b_{\rm min},\varpi_{\rm max}\tan b_{\rm min})$. As
shown in Fig.~\ref{cylmodel}, the resulting volume which will be used for our 
fit to the vertical profile is a double semi-infinite cylinder. For any
given star with $|b| > b_{\rm min}$, three regions are then defined: 
$0 < |z| < \varpi_{\rm max}\tan b_{\rm min}$, 
$\varpi_{\rm max}\tan b_{\rm min} \le |z| \le \varpi_{\rm max}\tan |b|$, and
$\varpi_{\rm max}\tan |b| < |z| < \infty$, and only the second one is used for
our calculation. A star with a large value of $\varepsilon_\pi$ will have 
non-negligible values of $f(z)$ in the three regions while the probability 
distribution for one with a small value of $\varepsilon_\pi$ will be
concentrated in one of the three (unless its value of $\pio$ is close to one of
the two critical values which correspond to the two boundaries 
between regions). In any case,
each of those stars will have an integrated probability in the second region
between 0 and 1, leading to a non-integral estimated number of stars in the
double semi-infinite cylinder. A third precaution to be considered is the 
introduction of biases in the obtained parameters by the fitting procedure. 
This problem can be solved using a twofold strategy: First, biases can be 
minimized using a variable bin size determined by the criteria specified by 
\citet{DAgoStep86}. Second, a population with the same parameters as those
derived in the fitting process can be generated numerically and fitted using the
same procedure. Repeating this a large number of times allows us to measure the
numerical biases by obtaining the mean displacements between the real and the 
derived parameters and to correct our results accordingly.

\section{THE SAMPLE}

	We have extracted from the Hipparcos catalog the data for all stars 
classified therein as O-B5 and with $|b| > b_{\rm min} = 5\degr$ (3\,531 
objects). That value of $b_{\rm min}$ was chosen in order to eliminate the 
highly obscured regions near the galactic plane while at the same time retaining
most of the O-B5 stars in the immediate solar neighborhood. 
Also, for $h\approx 45$ pc (a
typical value chosen from Table~\ref{previous}), a vertical distance of three 
scale heights corresponds to $\approx 1.5$ kpc in $\varpi$ for $|b| = 5\degr$, 
so we would not expect to sample regions with large variations in the 
early-type star population due to radial galactic gradients. From that 
sample we have selected a subsample (3\,382 objects) with good astrometric 
solutions by eliminating those stars with more than 5\% or data points 
rejected (column 29, \citealt{ESA97}) and with a goodness-of-fit statistic $g$
for $\nu$ degrees of freedom:

\begin{equation}
g = \left(\frac{9\nu}{2}\right)^{1/2}
    \left[\left(\frac{\chi^2}{\nu}\right)^{1/3} + \frac{2}{9\nu} - 1\right],
\label{ohmy}
\end{equation}

\noindent less than 3.0 (column 30)\footnote{The selected subsample is
available in electronic format at: 
{\tt http://www.stsci.edu/\~{}jmaiz/data/hipparcos}.}

	Since we expect most stars in our subsample to be located at distances 
$> 100$ pc and that for Hipparcos data, typically, $\sigma_\pi\approx 1$ mas, 
Lutz-Kelker-type biases should be important. Indeed, that is what can be 
deduced from Table~\ref{epsilondist}, where the distribution of 
$\varepsilon_\pi$ values is shown. 

\begin{deluxetable}{rr}
\tablecaption{Frequency of Hipparcos relative parallax errors for the selected 
subsample of 3\, 382 O-B5 stars used in this work. The last row refers to
stars with negative parallaxes.\label{epsilondist}}
\tabletypesize{\scriptsize}
\tablewidth{0pt}
\tablehead{\colhead{$\varepsilon_\pi$ range} & \colhead{Number}}
\startdata
$0.00-0.05$                  &      6 \\
$0.05-0.15$                  &    237 \\
$0.15-0.25$                  &    380 \\
$0.25-0.50$                  &    848 \\
$>0.50$                      & 1\,444 \\
\multicolumn{1}{l}{negative} &    467 \\
\enddata
\end{deluxetable}

	The other parameter which has to be selected to define the double 
semi-infinite cylinder which will be used in this work is $\varpi_{\rm max}$.
To do that, we have to consider that the Hipparcos completeness limit for 
early-type stars is determined by $m_V = 7.9 + 1.1|\sin b|$, which corresponds 
to $m_V = 8.0$ for $|b|=5\degr$. A very small fraction ($\ll 1\%$) of stars 
below the limit is actually not included and the catalog is partially complete 
between that limit and $m_V\approx 12-13$ \citep{ESA97}. A B5~V star has 
$M_V \approx -1.2$; at a distance of 350 pc such a star with $A_V = 1.0$ mag 
(obtained from a very conservative estimate of 3.0 mag/kpc in $V$)
would have $m_V = 7.5$, still within the completeness limit. If such a star
was located at 135 pc (approximately 3 scale heights) from the galactic plane
and with $\varpi = 350$ pc, its apparent magnitude would also be $\approx 7.5$
if we estimate $A_V$ as $\approx 0.5$ (again, a very conservative estimate at
that latitude). We can then conclude that a choice of 
$\varpi_{\rm max} = 350$ pc leads to a nearly complete sample for the O-B5 disk
stellar population. 

\section{RESULTS}

\begin{figure}
\centerline{\includegraphics*[width=\linewidth]{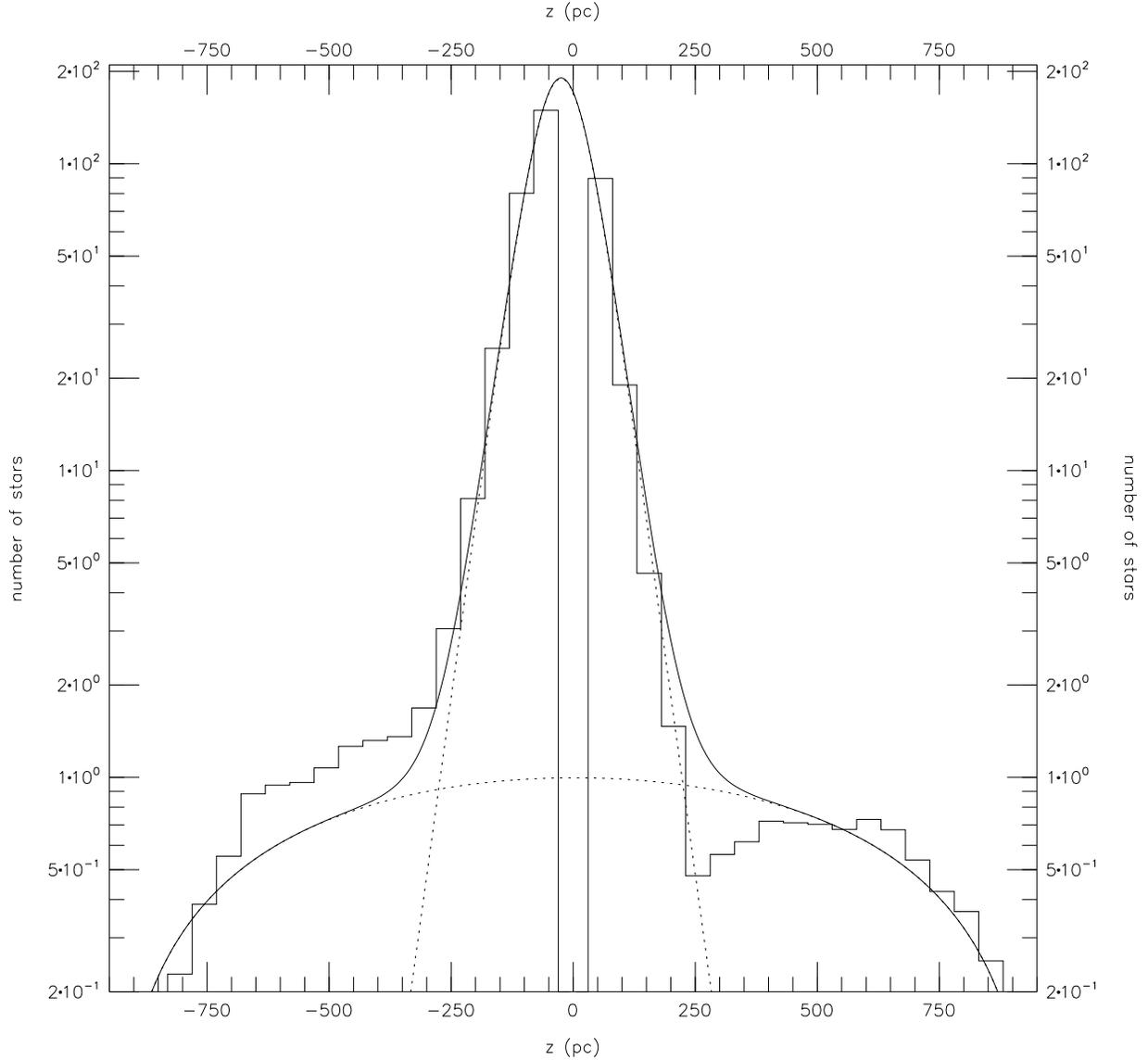}}
\caption{Observed (histogram) and model fit frequency distributions as a
function of the vertical coordinate for O-B5 stars using a self-gravitating
isothermal disk + parabolic halo distribution. The dotted lines represent each
individual component and the continuous one the sum of the two. The displayed
bin size is uniform and equal to 50 pc, but is not used for the fit (see text).
The region immediately surrounding $z=0$ is not considered for the fit, as it
corresponds to the space between the two semi-infinite cylinders. Note that the 
observed distribution shows ``fractional stars'' due to the procedure used to 
derive the histogram.}
\label{dists}
\end{figure}

\begin{figure}
\centerline{\includegraphics*[width=\linewidth]{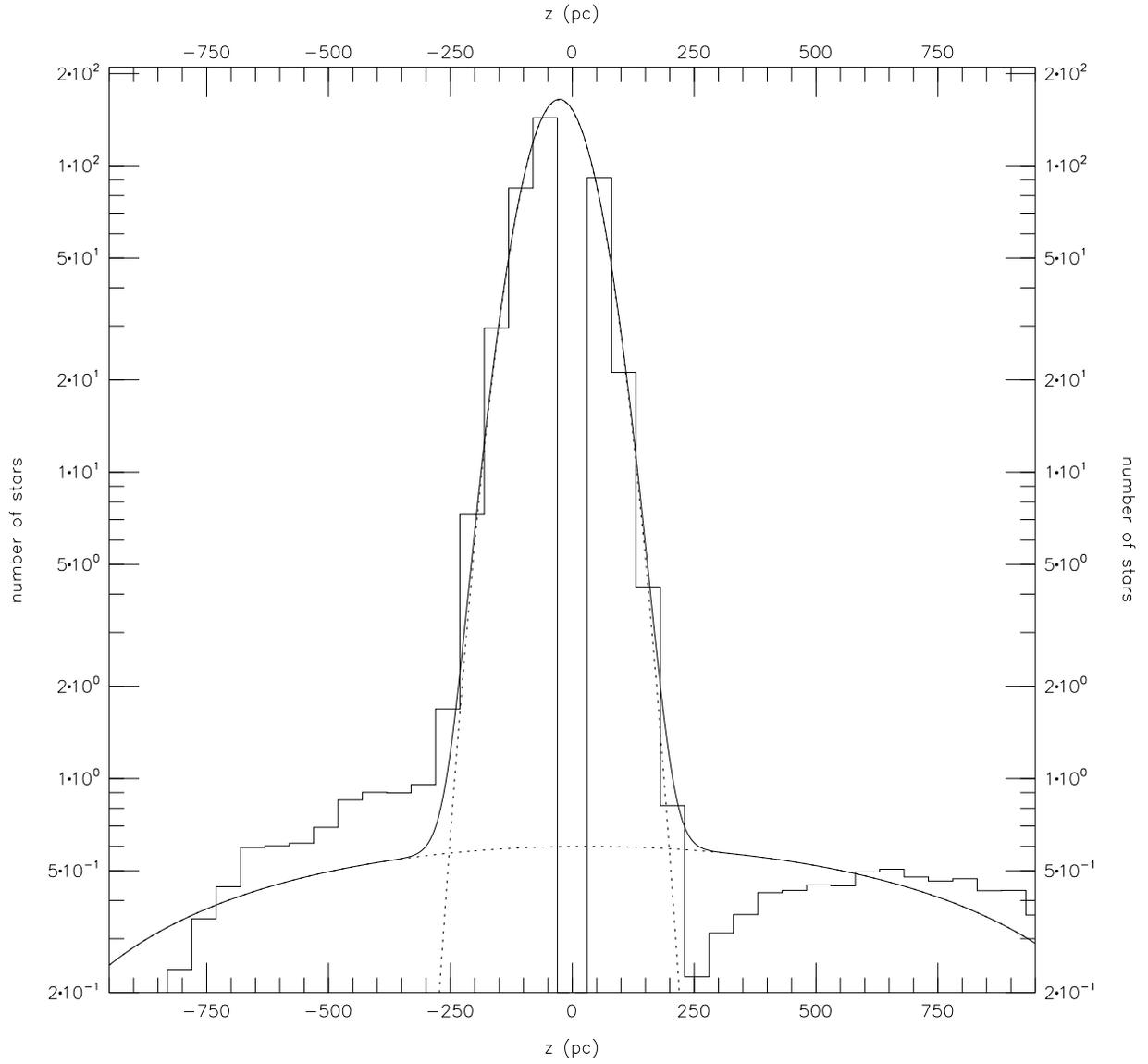}}
\caption{Same as Fig.~\ref{dists} but using a Gaussian disk model instead of
a self-gravitating isothermal one.}
\label{distg}
\end{figure}

	We have carried out least-squares fits to the data described in section 
3 using either a self-gravitating isothermal or a Gaussian disk model and 
applying the method described in section 2. Since the halo population in our
sample is clearly incomplete due to the Hipparcos magnitude limits, we used 
a simple reasonable model to describe it, a parabola. That choice allows us to
discriminate between the disk and halo contributions within a few hundred
parsecs of the Sun but can only produce a lower limit for \sigmastarh . The data
and fits are displayed in Figs.~\ref{dists}~and~\ref{distg} and the obtained
parameters are shown in Table~\ref{results}.

\begin{deluxetable}{cccccccc}
\tablecaption{Results of the least-squares fits using either a self-gravitating
isothermal (first row) or a Gaussian (second row) disk + parabolic halo 
distribution. The values have been bias-corrected, and both the random (first)
and systematic (second) uncertainties are shown in each case. The $h$ column
refers to the scale height ($h_s$) in the first row and to the half-width
($h_g$) in the second row. No uncertainties are given for 
$\sigmastarh/\sigmastard$ because the fitted values are only lower-limits to the
real distributions. \label{results}}
\tabletypesize{\scriptsize}
\tablewidth{0pt}
\tablehead{Model & $h$  & \zsun/ & \sigmastard                 & 
$\sigmastarh/\sigmastard$ & $\chi^2/\nu$ & $\nu$ & $g$ \\
                 & (pc) & (pc)   & (10\mmm/ O-B5 stars pc\mm/) &
                          &              &       &     }
\startdata
1 & $34.2\pm 0.8\pm 2.5$ & $24.2\pm 1.7\pm 0.4$ & $1.62\pm 0.04\pm 0.14$ &
0.058 & 0.88 & 41 & $-0.50$ \\
2 & $62.8\pm 1.7\pm 4.7$ & $25.2\pm 1.5\pm 0.5$ & $1.67\pm 0.04\pm 0.24$ &
0.037 & 0.87 & 41 & $-0.56$ \\
\enddata
\end{deluxetable}

	As previously discussed, a variable bin size was used in order to
minimize numerical biases. For each of the two models we generated 50
populations with the same parameters derived from the real population and we
estimated the biases by calculating the difference between the mean values of
the derived parameters and the real ones. As expected, those biases where very
small, typically 0.5 pc for $h$ and \zsun/. However, one possible bias due to
extinction still remains. Suppose we are observing a star with $b=-5\degr$
located inside the double semi-infinite cylinder but with a high value of
$\varepsilon_\pi$: This star will be counted only as a ``fractional star'' (i.e.
its integrated probability inside the volume of interest will not be close to
one). If the sample were complete both inside and outside the double
semi-infinite cylinder this would not be a problem, since the lost probability
would be compensated by that contributed by stars located outside the cylinder.
However, we expect that extinction (and, to a lesser degree, distance) will
remove from our sample some of the stars outside the cylinder, especially those
closer to the galactic plane. To simulate this effect we consider an extreme
case by placing a wall of infinite extinction at $\varpi=\varpi_{\rm max}$
which extends one scale height (for the self-gravitating isothermal case) or
one half-width (for the Gaussian case) north and south of the galactic plane. 
The volume excluded by this wall is represented by the hatched area 
in Fig.~\ref{cylmodel} for the self-gravitating isothermal case. Note that only 
stars in the southern galactic hemisphere are eliminated by this wall since the 
position of the Sun above the galactic plane locates any star with $b=5\degr$ 
above the wall\footnote{For the Gaussian case a small volume is also eliminated
in the northern galactic hemisphere.}. Under these conditions, we eliminated 
the stars in the hatched volume from our generated 50 populations and calculated
the corresponding biases by calculating new least-squares fits. As expected, 
the new biases turned out to be more relevant than the ones obtained before, 
but even then they never induced changes larger than 15\% in the values of the 
parameters.  Since the real situation must be in between the two extreme cases 
discussed here ((a) all stars in the possibly excluded area partially 
contributing to the probability inside the cylinder and (b) none of them doing 
so), we considered the real bias to be the average of the two and we indicated 
the two extremes as a systematic error. 

	The results in Table~\ref{results} indicate that both the
self-gravitating isothermal and the Gaussian models are very good descriptions 
of the disk population, since $g$ is close to 0 in both cases and the fitted
distribution near the galactic plane closely resembles the observed one, as
shown in Figs.~\ref{dists}~and~\ref{distg}. The parabolic fit to the halo
population is of worse quality, as expected from the much smaller number of
stars in our sample belonging to it and from its incompleteness. Nevertheless,
the existence of a halo population in our sample is beyond doubt, as is clearly
shown in Figs.~\ref{dists}~and~\ref{distg}. It is also well known that this halo
population extends farther than the distance-induced limits in our sample, even
beyond a few kpc \citep{Rolletal99}. Thus, the values of
$\sigmastarh/\sigmastard$ which appear there have to be understood only as lower
limits to the real ones. 

	The values of \zsun/ and \sigmastard\ obtained using the two different
models are clearly compatible. Since we are unable to determine which of the two
models provides a better fit to the disk population, either result can be used.
Probably the ones obtained with the self-gravitating isothermal profile are
preferred due to the better quality of the parabolic fit to the halo
population, which should introduce less numerical noise into the result than 
in the Gaussian case.

	Our value for $h_s$ is somewhat lower than the ones obtained by other
authors. The discrepancies are only at the $2\sigma$ level with the values 
obtained by \citet{ContVacc90} and by \citet{StotFrog74} (within 200 pc of the 
Sun) and even lower in comparison with the \citet{Reed00} result. 
\citet{Alle73} does not 
quote an uncertainty while the value of \citet{StotFrog74} within 800 pc of the 
Sun is quite probably affected by heavy extinction. Some of the discrepancies
may be attributed to contamination by halo stars (a non-negligible effect, as
we have seen) and others to the use of a purely exponential model instead of a 
self-gravitating isothermal one. Our value of \zsun/ is clearly compatible with 
the one obtained by \citet{StotFrog74} but is higher than the ones obtained by 
\citet{ContVacc90} and \citet{Reed97}, in this last case by a large amount.
We believe that the differences may be caused by the difficulty in accounting
for complex extinction variations at low galactic latitudes, especially when 
one has to rely on spectroscopic parallaxes, a problem which does not affect our
results since they are based on unbiased trigonometric distances. This can be
seen in the fact that the result more similar to ours is the one which is
obtained from the nearest population. Finally, our value for $\sigma_\star$ is
within the range obtained by \citet{Reed00} but is clearly higher than the one
obtained by \citet{StotFrog74} and lower than the one obtained by
\citet{Alle73}. Once again, we believe that extinction may be responsible for
some of the discrepancies but we must also take into account that $\sigma_\star$
is expected to suffer significant variations depending on the maximum radius
used, since different regions of the disk (close or far from spiral arms) are
expected to be sampled.

	One last point to discuss is the existence of spatial variations within
the region discussed in this paper. \citet{StotFrog74} discovered that a 100 pc
radius hole exists in the distribution of O-B5 stars at low galactic latitudes. 
Also, it has been known for a long time that a large fraction of the early-type 
stars in the solar neighborhood belong to Gould's belt, a structure inclined 
$15-20\degr$ with respect to the galactic plane and which is expanding with 
respect to a point located close to the Sun's position (see \citealt{Torretal00}
for a recent analysis). Is this expected minimum in the number density of
early-type stars detected in the Hipparcos data? It indeed is. The parameters 
derived from our self-gravitating isothermal model predict that there should be
11.6 O-B5 stars within 66.67 pc of the Sun (i.e., with $\pi \ge 15$ mas). 
However, only three\footnote{The three have small values of $\varepsilon_\pi$ 
so their position can be determined with certainty to be within that volume. Two
other stars (HIP 23342 and HIP 74368) have $\pio \ge 15$ mas but with large
values of $\varepsilon_\pi$ and with spectroscopic parallaxes which clearly
indicate that they are much farther away. Finally, a few stars have values of
$\pio$ slightly below 15 mas and could actually be closer to us than 66.67 pc.
However, the probabilistic estimation yields a number lower than one, and even 
if there was an additional star within the given distance our arguments would
remain unchanged.} O-B5 stars are found in that volume: $\eta$~UMa,
$\alpha$~Eri, and $\alpha$~Pav. The difference between the expected and the
measured number appears to be too large to be explained as a statistical
fluctuation, so we conclude that the existence of a local ``hole'' in the
distribution of early-type stars is indeed real.

	We conclude that our method provides a solid measurement of the
vertical distribution of O-B5 stars in the solar neighborhood by using
trigonometric parallaxes alone and by taking into account all relevant biases.

\acknowledgments

The author would like to thank Ivan King for introducing him some years ago 
to the problems caused by the Lutz-Kelker bias and Stefano Casertano for very
fruitful discussions regarding the topics in this article. Useful comments for 
improving this paper were also provided by an anonymous referee and by
Nolan R. Walborn. Support for this work was provided by NASA through grant 
GO-8163.01-97A from the Space Telescope Science Institute, Inc., under NASA 
contract NAS5-26555, and by the STScI DDRF.

\bibliographystyle{aj}
\bibliography{general}

\end{document}